 \documentclass[aps,amssymb,a4paper,amsmath,twocolumn  ] {revtex4}%twocolumn
\usepackage{graphicx,subfigure}% Include figure files
\usepackage{bm}
\usepackage{amsmath,mathrsfs}
\usepackage{amssymb}
\usepackage{slashed}
\usepackage{color}
\usepackage{ulem}
\usepackage{amsfonts}
\usepackage{amsmath}
\usepackage{graphicx}

\begin{document}
	%	\begin{CJK}{GBK} {song}
	\title{ 
		$ \rho\to \pi\pi $ Hadronic Decay in the Nambu-Jona-Lasinio Model: Mass-Width Interplay and Beyond-RPA Corrections
		
		  }
	
	\author{Qing-Wu Wang    }~\email[]{Email:  qw.wang@scu.edu.cn  } %orcid:0000-0001-5458-8135
	\author{Xiao-Fu \L\text{\"{u}}    }
	\author{Hua-Zhong Guo    } ~\email[]{Email:  guohuazhong@scu.edu.cn  }%orcid:0000-0003-0478-8475

	\affiliation{
		 College of Physics, Sichuan University,  Chengdu 610064, China}
	
	%%%%%%%%%%%%%%%%%%%%
	\begin{abstract}

	 We present a novel framework for analyzing unstable composite particles using Green's functions and dispersion relations. As an illustrative example, we explore the
	 $\rho$ vector meson decay process $\rho\to\pi\pi$
	 within the Nambu -- Jona - Lasinio (NJL) model. Our approach addresses a key limitation of the four-quark interaction description, which adequately describes two-quark bound states but fails to describe decay
     processes. The Bethe-Salpeter(BS) wave function of the $\rho$ meson exhibits time evolution that leads to the physical mass
	 $M$ incorporating a correction $\Delta M $. This correction depends on the decay width  $\Gamma(M)$. This work provides crucial insights into the dynamical relationship between resonance masses and their decay properties, addressing a long-standing challenge in hadron physics. The calculated mass and width are in good agreement with the experimental values, demonstrating the effectiveness of this approach for studying unstable hadronic systems beyond conventional bound-state approximations.

%	 $\mathbf{Keywords:}$ Strong decay, Bethe-Salpeter  equation, Prepare state, Mass correction
	
	\end{abstract}

	\maketitle
		
	\section{Introduction}

	Over the last few decades, we have witnessed significant advancements in experimental facilities such as Belle, BES, CLEO, CDF, LHC, and BABAR, generating vast amounts of data and revealing unexpected discoveries in hadronic system studies \cite{Olive_PDG_2014,Belle_CPV, BESIII_Zc,CLEO_Decay, CDF_Top, BABAR_sin2b,ATLAS_CMS_Higgs}.
		Understanding meson properties, particularly their masses and decay dynamics, is crucial for probing the nonperturbative regime of quantum chromodynamics (QCD).	
	As QCD remains analytically intractable at low energies, bound state systems have long been studied using various methods and models\cite{Roberts_2021,Shifman_2009,Brodsky_2015, Klevansky_2019, Fischer_2018, Maris_2003,  Wang2018}.	
 These approaches provide insights into the mechanisms of chiral symmetry breaking, hadron structure, and the interplay between bound states and decay channels.  Recent advances in high-energy physics experiments have uncovered numerous exotic resonance states, challenging traditional decay theories based on static bound states.
	However, persistent challenges still lies in accurately predicting meson masses, especially for unstable systems such as resonances, where higher-order corrections and decay dynamics significantly influence physical observables.
	
		Effective models such as the Nambu--Jona-Lasinio (NJL) model and the Bethe-Salpeter (BS) equation formalism have emerged as powerful tools to bridge the gap between fundamental theory and observable phenomena. The NJL model, rooted in spontaneous chiral symmetry breaking, describes mesons as quark-antiquark bound states through a four-fermion interaction. Within this framework, the random phase approximation (RPA) offers a leading-order description of meson masses and couplings.
		 For instance, the RPA predicts a $\rho$ meson mass of 834 MeV, which overestimates the empirical value of (770 MeV).
		 Traditional homogeneous BS equations inherently describe stable bound states while neglect decay-induced mass shifts. 	Incorporating Self-energy corrections through $1/N_c$-expansion technique systematically account for higher-order processes,lowering the $\rho$ mass and aligning theoretical predictions with experimental data, which demonstrates the importance of beyond-RPA contributions in refining mass calculations   \cite{Polleri9611, Oertel2000}.
		
		  %The result of decay width $\Gamma$ is 118 MeV which is improved from earlier calculation with $\Gamma=$\cite{klimt1990}.	
	
%Complementary to quark-level models, the BS equation provides a relativistic framework for studying mesons as composite systems. For exotic resonances like $\chi_{c0}(3915)$, interpreted as meson-meson molecular states, the BS formalism captures the internal structure through wave functions derived from meson-meson interactions.

	 However, recent works has demonstrated that unstable states cannot be treated as static bound states\cite{Chen2022, Chen2023}. The crucial advancement lies in extending BS theory to include unstable systems by integrating time evolution and dispersion relations. Evaluating $T$-matrix elements across open and closed decay channels (e.g., $J/\psi \omega$ and $D\bar{D}$) reveals mass corrections arising from principal value integrals over final-state energy continua. For $\chi_{c0}(3915)$, this approach yields a physical mass of 3922 MeV, consistent with experiments, and highlights the significant role of decay channels like $J/\psi \omega$ in reducing the bare molecular-state mass (3954 MeV) by ~30 MeV. Such refinements underscore the necessity of coupling bound-state dynamics to decay processes in resonance studies and it provides a feasible scheme to describe meson decay.
	
%	Despite these advances, challenges remain in reconciling model parameters (e.g., cutoff scales, coupling constants) with QCD constraints and experimental data.	
	% The NJL model’s reliance on regularization schemes and the BS approach’s dependence on phenomenological meson-quark couplings introduce uncertainties. Moreover, the treatment of confinement and the convergence of perturbative expansions (e.g., $1/N_c$) demand further scrutiny. Future work must integrate these frameworks with lattice QCD results and experimental inputs to refine predictive power, particularly for newly observed exotic states. By addressing these challenges, theoretical models will deepen our understanding of hadron structure and the emergent phenomena of QCD in the nonperturbative domain.

 Unlike traditional models where bound state masses are fixed and decay widths are externally imposed parameters, our approach self-consistently determines masses and widths through system evolution, eliminating independent assumptions about these quantities.
 Thus the total Hamiltonian is split into two components: one constructs the initial bound state (``prepared state") via the homogeneous BS equation, and the other introduces perturbative interactions driving decay. The prepared state represents a metastable configuration formed under strong interaction equilibrium, while decay occurs through coupling to continuum states. %For example, a quark-antiquark pair transitions to lower-energy states via meson exchanges before decaying into stable particles.
  The decay dynamics are characterized by using the analytic properties of Green's functions, where complex-plane poles directly link to mass and width. Different from traditional methods treating width as a function of mass, our approach reveals mutual dependence through analytic continuation. It resembles $  1/N_c  $ approximations \cite{Polleri9611} but involves distinct mechanisms.
Decay interactions are not mere energy corrections but drivers of system evolution.% For instance, light vector meson  exchanges generate attractive forces in charmonium decays, forming near-threshold resonances. Insufficient interaction strength leads to finite lifetimes due to centrifugal barriers, rather than full binding.   %\cite{Feldmann2000,du2021}.

 In this  paper, we follow the idea in Ref. \cite{Chen2023}, treating resonance  as a non-stationary bound states. The time evolution of two-body bound state is determined by the total Hamiltonian. According to dispersion relation, the total matrix elements for all
 decay channels are calculated with respect to arbitrary value of the final state energy,
 and these matrix elements are expressed in terms of the Heisenberg picture.
 The mass correction includes all decay final states.
 Since the $\rho$ meson decay is dominated by $\rho\to\pi\pi$ channel, we use this method to study the $\rho$ meson decay process,simplifying the calculation.
  This decay channel is chosen because its decay width is neither too small (ensuring that the mass correction cannot be ignored) nor too large, allowing for approximate treatment.
 In high-energy physics experiments, the
 $ \rho $  meson and its decay to $\pi\pi$
 can be used as a reference process. The decay signature of two pions can be used to calibrate detectors, as well as to study the background characterization. In high-energy collider experiments such as those at the Large Hadron Collider (LHC), the reconstruction of $\rho$ meson decays provides a valuable benchmark for verifying detector performance and optimizing signal-to-noise discrimination. Furthermore, this well-understood decay channel offers a crucial reference process for searches of new physics phenomena. \cite{Sibirtsev:1997ac,Rebyakova2012}.

  The paper is organized as follows: Sec.\ref{sec.njl} presents the formlism of the NJL model, and then give $ \rho \to \pi \pi   $ meson  decay results based on the framework that is proposed in Ref. \cite{Chen2023} in Sec. \ref{sec.decay}, and Sec. \ref{sec.sum} provides concluding remarks. Some technical details of the calculations are collected in Appendix \ref{sec.fulu}.

 %\section{Formula and results}

  \section{NJL model}

 \label{sec.njl}

 The Nambu–Jona-Lasinio (NJL) model provides an effective field-theoretic framework for studying the properties of hadrons, particularly mesons, in the context of low-energy quantum chromodynamics (QCD). The model is based on the idea of spontaneous chiral symmetry breaking, which is a key feature of QCD, and it provides a framework for understanding the structure and dynamics of mesons as bound states of quarks and antiquarks \cite{Buballa2005,klimt1990,Bernard1993,Klevansky:1992qe}.
 Nevertheless, the NJL model's inability to incorporate color confinement creates difficulties in handling unphysical quark-antiquark thresholds, requiring additional phenomenological constraints to supress Landau damping effects.

  Within the NJL model, meson modes are viewed as collective
 solutions of the Bethe-Salpeter equation (BSE) in the framework of the random phase approximation (RPA).  At least at the RPA level, mesons seem to emerge as poles in the corresponding
$  T $ -matrix channel.
 The NJL model is characterized by a four-fermion interaction Lagrangian, which describes the interactions between quarks. For the case of two flavors (up and down quarks), the Lagrangian is given by:
\begin{eqnarray}
	 \mathcal{L}_{\text{NJL}} = \bar{\psi}(x) \left(i\slashed{\partial} - m_0\right) \psi(x) + \mathcal{L}_{\text{int}},
\end{eqnarray}
 where $\psi(x)$ represents the quark field, $m_0$ is the bare quark mass, and $\mathcal{L}_{\text{int}}$ is the interaction term, it can be typically written as:
\begin{eqnarray}
	 \mathcal{L}_{\text{int}} = \frac{G_1}{2} \left[ \left(\bar{\psi}(x)\psi(x)\right)^2 + \left(\bar{\psi}(x)\gamma_5 \tau^a \psi(x)\right)^2 \right] + \nonumber
\\
	 \frac{G_2}{2} \left[ \left(\bar{\psi}(x)\gamma_\mu \tau^a \psi(x)\right)^2 + \left(\bar{\psi}(x)\gamma_\mu \gamma_5 \tau^a \psi(x)\right)^2 \right].
\end{eqnarray}
 Here, $G_1$ and $G_2$ are coupling constants for the scalar-pseudoscalar and vector-axial vector channels, respectively, and $\tau^a$ are the Pauli matrices representing isospin.

 %Quark Self-Energy and Gap Equation

 In the NJL model, the quark self-energy (or constituent quark mass) is generated dynamically through the gap equation. The gap equation within the Hartree approximation is given by:

\begin{eqnarray}
	 m_q = m_0 + m_q G_1 N_c 8i \int^\Lambda \frac{d^4p}{(2\pi)^4} \frac{1}{p^2 - m_q^2},
\end{eqnarray}
 where $m_q$ is the dynamically generated quark mass, $N_c$ is the number of colors, and $\Lambda$ is a momentum cutoff introduced to regularize the divergent integral. The gap equation describes how the quark mass is modified due to the strong interaction, leading to the phenomenon of spontaneous chiral symmetry breaking.

% Bethe-Salpeter Equation for the T-Matrix

 Mesons can be treated as bound states of quarks and antiquarks,  their properties are determined by solving the BSE. The BSE for the $ T $-matrix in the  RPA   is given by:
\begin{eqnarray}
	 i {\mathcal{T}}(q^2) &=& i\mathcal{K} - \text{Tr} \int \frac{d^4p}{(2\pi)^4}  i\mathcal{K} iS(p+\tfrac{1}{2}q)\nonumber \\
& &	\times i {\mathcal{T}}(q^2) iS(p-\tfrac{1}{2}q),
\end{eqnarray}
 where $\mathcal{K}$ is the interaction kernel, and $S(p)$ is the quark propagator. The $ T $-matrix can be decomposed into different channels corresponding to scalar, pseudoscalar, vector, and axial vector mesons.

 We focus on the application of the NJL model to the study of the rho meson ($\rho$), as discussed in  Refs. \cite{Polleri9611,Xiao2013Heavy,V1988Properties,C1993Vector}. For the vector meson channel (e.g., the rho meson), the T-matrix takes the form:

\begin{eqnarray} \label{eq.rmass}
	 i\mathcal{T}_{\rho}(q^2) = (\gamma_\mu \tau^a \otimes \gamma_\nu \tau^a) \frac{G_2}{1 - G_2 J_{VV}(q^2)} \left(g^{\mu\nu} - \frac{q^\mu q^\nu}{q^2}\right),
\end{eqnarray}
 where $J_{VV}(q^2)$ is the polarization function for the vector channel.  An explicit expression for this  function  can be found in  Ref. \cite{klimt1990} where  $J_{VV}=\frac{2}{3}[(2m^2+q^2)I(q)-2m^2I(q^2=0)]$ and   $ I(q) $ is defined in the appendix \ref{sec.fulu}.
 The masses of the bound states are obtained from the position of the poles  of the T-matrix.

%Rho Meson Decay $\rho \to \pi\pi$

 The decay width $\Gamma_{\rho \to \pi\pi}$ is calculated by evaluating the leading-order contribution to the decay amplitude. It is given by:
\begin{eqnarray}\label{eq.width}
	 \Gamma_{\rho \to \pi\pi} = \frac{N_c^2}{12\pi} g_{\rho q\bar{q}}^2 g_{\pi q\bar{q}}^4 m_\rho \left(1 - \frac{4m_\pi^2}{m_\rho^2}\right)^{3/2} G^2(m_\rho^2; m_\pi^2),
\end{eqnarray}
 where $g_{\rho q\bar{q}}$ and $g_{\pi q\bar{q}}$ are the coupling constants of the rho and pion to the quark-antiquark pair, respectively,  which can be defined  in terms of the functions $ J_{VV} $, $ J_{PA}  $ and $ J_{ AA } $.
And $G(m_\rho^2; m_\pi^2)$ is a form factor that depends on the momentum transfer.  %As the strong interaction is described by QCD, and the $\rho$ meson decay provides a laboratory to test QCD predictions at relatively low energies.

 %The calculated decay width in the NJL model is found to be in reasonable agreement with the experimental value, demonstrating the model's ability to describe meson decays.

 \begin{figure}
 	\centering
 	\includegraphics[width=0.7\linewidth]{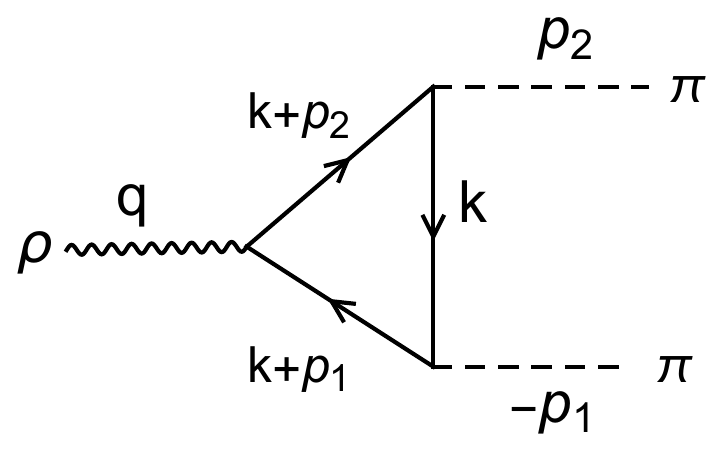}
 	\caption{ The  amplitude used to calculate the decay width. With the  amplitude $ (-p_1\leftrightarrow p_2) $, it corresponds to the function
 		$ T^\mu (q^{2};p_{2}^{2},p_{1}^{2})) $ as in Eq.\eqref{eq.t0}.}
 	\label{fig.decay}
 \end{figure}

  \section{ The decay of $\rho\to\pi \pi$ and the prepared state  }

 \label{sec.decay}

 In the De R\'{u}jula-Georgi-Glashow description of hadrons, the $\rho$ mesons, as a bound state of a quark and an anti-quark, is interpreted as an excited version of the pion.
 It has a spin-parity of $ J^P=1^- $. The $\rho$ meson predominantly decays into two pions via the strong interactions. The decay width $\Gamma$ of the $\rho$ meson is relatively large due to the strong interactions nature of the decay. The branching ratio for
 $ \rho \to \pi \pi  $ is very high, close to 100\% \cite{PDG2024}.  Examples of a calculation of the $\rho$ meson decay
 width and $\pi\pi$ scattering phase shifts from lattice QCD can be found in Refs. \cite{Brice2018,Bulava2016,Rodas2023}.
  For the
$ \rho_0 $, its total decay width is about 149 MeV, and almost all of it decays into  $  \pi^{ +} \pi^{-} $ while
$ \pi^0\pi^0 $ is forbidden. This high branching ratio is a characteristic feature of the
$ \rho \to \pi \pi  $ decay mode which is a prime example of a strong interaction process. In the rest frame of the
$ \rho $ meson, the two pions are emitted with specific momenta and the leading contribution to
this process corresponds to the diagram shown in Fig.  \ref{fig.decay}.

  Effective  meson-$ \bar q q  $ vertices is obtained  within  the  RPA  approximation
 \begin{align}
 	i\Gamma^{\rho}&=ig_{\rho - q\bar{q}}\gamma^{\mu}\tau^{a} \end{align}
 and
  \begin{align}
 		 i\Gamma_{\pi}=ig_{\pi - q\bar{q}}(1 - a_{\pi}\hat{\slashed{k} } )\gamma^{5}\tau^{a},
 \end{align}
with $ \hat{\slashed{k} }= \slashed{k} /k^2   $.
From Fig. \ref{fig.decay}, one obtains the amplitude for the decay process by using the usual Feynman rules
\begin{eqnarray}
		\mathcal{M}(q;p_1,p_2) =\epsilon_{\mu}(q)T^{\mu}(q;p_1,p_2),
\end{eqnarray}
where
\begin{eqnarray}\label{eq.t0}	
	T^{\mu}(q;p_1,p_2)&=&-tr\int\frac{d^{4}k}{(2\pi)^{4}}ig_{\rho - q\bar{q}}\gamma^{\mu}\frac{i}{\slashed k+\slashed p_2 - m_q}  \nonumber\\
	&&\times i\sqrt{2}g_{\pi - q\bar{q}}(1 + a_{\pi}\hat{\slashed{p}}_2)i\gamma^{5}\frac{i}{\slashed k - m_q}\nonumber\\
	&&\times i\sqrt{2}g_{\pi - q\bar{q}}(1 - a_{\pi}\hat{\slashed{p }}_1)\nonumber\\
&&\times	i\gamma^{5}\frac{i}{\slashed k+\slashed p_1 - m_q}-(-p_1\leftrightarrow p_2).
\end{eqnarray}
The total momentum of the $\rho$ meson is $  q =p_2 - p_1. $ The negative sign in front of the momentum
$ p_1 $ makes the quark propagator possess symmetry under the exchange of $ p_1 $ and $ p_2 $. This is not essential, but it simplifies the subsequent handling of the form factor of that decay.

 Evaluating the trace one gets
\begin{eqnarray}\label{eq.t12}
	T^{\mu}(q;p_1,p_2)&=&-2iN_cg_{\rho - q\bar{q}}g_{\pi - q\bar{q}}^{2}\times \nonumber\\
&&[p_{1}^{\mu}G(q^{2};p_{1}^{2},p_{2}^{2})+p_{2}^{\mu}G(q^{2};p_{2}^{2},p_{1}^{2})].~~~~~~~~~~
\end{eqnarray}
 For the decay $  \rho \to \pi \pi $, the two pions are on shell and we have $ p_{1}^{2}=p_{2}^{2} $. Then   $ T^\mu $ can be expressed as
 	 \begin{eqnarray}\label{eq.t}
 		 T^{\mu}(m_\rho;m_\pi) =-2iN_cg_{\rho - q\bar{q}}g_{\pi - q\bar{q}}^{2}
   ( p_{1} ^{\mu}+p_{2} ^{\mu})G(m_\rho^{2};m_\pi^{2}) ~~~
 	\end{eqnarray}
   with   $ G(m_\rho^{2};m_\pi^{2})  $  the form factor as in Eq.\ref{eq.width}. The calculation of  $ G(m_\rho^{2};m_\pi^{2})  $  can be found in the Appendix. As the form factor $ G $ is obtained, the decay width is available from Eq. \eqref{eq.width} which depends on the $\rho$ meson mass. The mass of $\rho$ meson defined through the pole of Eq.\eqref{eq.rmass} is independent of its decay width. In many cases, it is a good approximation as the decay width is not large enough. But for most molecular bound states and here the $\rho$ meson, the decay width gives a small contribution to the physical mass. %However, the decay widths are not described by a Breit-Wigner form.
   In Ref.  \cite{ke1965}  from the requirements of analyticity and unitarity, the resonance behaviours of the partial wave amplitudes for $\pi\pi$ scattering  involves a function deviating from $ 1 $, which indicates deviation from the Breit-Wigner type resonance formula.

    In Ref. \cite{Polleri9611}, the self-energy correction to the $\rho$ meson mass beyond the RPA approximation is calculated at order $1/N_c$. The self-energy $\mathbf{\Pi}^{\mu\nu}(q)$ is decomposed into real ($\mathcal{R}e\,\mathbf{\Sigma}$) and imaginary ($\mathcal{I}m\,\mathbf{\Sigma}$) parts \cite{Polleri9611}. The imaginary part, derived using cutting rules, arises from pion-loop contributions   and is expressed as:
    \begin{eqnarray}
    \mathcal{I}m\,\mathbf{\Sigma}(q^2) = -q\,\Gamma_\rho(q),
   \end{eqnarray}
    where $\Gamma_\rho(q)$ relates to the $\rho \to \pi\pi$ decay width. The real part is obtained via the Kramers-Kronig relation:
    \begin{eqnarray}
    \mathcal{R}e\,\mathbf{\Sigma}(q^2) = \frac{1}{\pi}\mathcal{P} \int_{4m_\pi^2}^{\Lambda^2} \frac{\mathcal{I}m\,\mathbf{\Sigma}(\mu^2)}{\mu^2 - q^2} d\mu^2,
    \end{eqnarray}
    with a consistent cutoff $\Lambda$ applied to all loops. Including $1/N_c$ corrections, the rho mass is shifted from its RPA value ($m_\rho^{(0)}$) by solving:
    \begin{eqnarray}
    m_\rho^2 = \left(m_\rho^{(0)}\right)^2 + \mathcal{R}e\,\mathbf{\Sigma} \left(m_\rho^{(0)}\right)^2 ,
    \end{eqnarray}
    Using model parameters, this yields a physical mass of $m_\rho = 770\,\text{MeV}$, consistent with experimental data, and gives the decay width $\Gamma=118$ MeV   smaller than the experimental value. The imaginary part governs the decay width, while the real part renormalizes the mass. The result justifies the use of  Rayleigh-Schrodinger (RS) perturbation theory, contrasting with the   method that requires solving $m_\rho^2 = \left(m_\rho^{(0)}\right)^2 + \mathcal{R}e\,\mathbf{\Sigma}(m_\rho^2)$.

    Disregarding the influence of self-energy corrections on the mass of mesons, we regard the BS bound state as a preparation state that is ready to decay at any time. Only this ``prepared state" can be described by the ground-state BS wave function at $t=0$. The time evolution of the prepared unstable state $\mathscr{X}_a^{\rm ps}$ is
        described by    using the Green's function $G(\varepsilon) = (\varepsilon - H)^{-1}$  with
\begin{eqnarray}
	\mathscr{X}(t) = e^{-iHt}\mathscr{X}_a^{ps}=\frac{1}{2\pi i} \int_{C_2} d\varepsilon \, e^{-i\varepsilon t} \frac{1}{\varepsilon - H} \mathscr{X}_a^{\rm ps},
\end{eqnarray}
 where   the total Hamiltonian $H=K_I+V_I$. Here, $K_I$  and $V_I $ are   the interaction responsible for the formation of stationary bound state
 and  for the decay of resonance, respectively.
 The  subscript $ a $ represents various   final states.
 The contour $C_2$ ensures analytic continuation in the energy plane, avoiding singularities of $G(\varepsilon)$ \cite{Goldberger1964}.

  At $ t>0 $, the  probability amplitude of finding the time-dependent system $\mathscr X(t)$ in the prepared state $\mathscr X_a^{ps}$ is
\begin{eqnarray}
	 {\mathcal A}_{a}=( {\mathscr X}_{a}^{\mathrm{p s}}, {\mathscr X} ( t ) )={\frac{1} {2 \pi i}} \int_{C_{2}} d \epsilon{\frac{e^{-i \epsilon t}} {\epsilon-M_{0}-T_{a a} ( \epsilon)}}.
\end{eqnarray}
Compared with Ref. \cite{Chen2023}, the coefficient $ (2\pi)^3 $ in front of the $ T $-matrix is missing, but it is consistent with Ref. \cite{Goldberger1964}. This is due to the different normalization treatments of the BS wave function.
Here, the Green's function
\begin{eqnarray}
	G_{aa}(\varepsilon) = \frac{1}{\varepsilon - M_0 -   T_{aa}(\varepsilon)} 
\end{eqnarray}
 is related   to the $T$-matrix element $T_{aa}(\varepsilon)$, where $M_0$ is the bare mass of the bound state. The $ T $-matrix element is defined through
\begin{eqnarray}
	 \langle a \ \mathrm{o u t} | a \ \mathrm{i n} \rangle&=&\langle a \ \mathrm{i n} | a \ \mathrm{i n} \rangle \nonumber \\
	 &-&i ( 2 \pi)^{4} \delta^{( 4 )} ( P-P_a ) T_{a a} ( \epsilon).
\end{eqnarray}
 This formula connects the bound-state dynamics to scattering processes.
Consider the analyticity and unitarity of $ T $-matrix element, we can decomposes the $T$-matrix into real ($\mathbb{D}$) and imaginary ($\mathbb{I}$) parts on the second Riemann sheet with
\begin{eqnarray}
	T_{aa}(\varepsilon) = \mathbb{D}(\varepsilon) - i\mathbb{I}(\varepsilon).
\end{eqnarray}
The imaginary part $\mathbb{I}(\varepsilon)$ governs decay processes, while $\mathbb{D}(\varepsilon)$ contributes to mass renormalization. Considering only one decay channel, the  imaginary part can be expressed as
$  \mathbb{I}(\varepsilon) \sim \sum_b \delta^{(3)}(\mathbf{P}_b - \mathbf{P}) \delta(E_b - \varepsilon) |T_{ba}(\varepsilon)|^2 $,
where the sum is over final states $b$ and $\varepsilon$ is the total energy of final state. The delta functions enforce energy-momentum conservation, and $|T_{ba}|^2$ quantifies transition probabilities to decay channels.

The $ T$-matrix element obeys the dispersion relation, which gives

\begin{eqnarray}
	 \mathbb{D}(\varepsilon) = -\frac{\mathcal{P}}{\pi} \int_{\varepsilon_M}^\infty \frac{\mathbb{I}(\varepsilon')}{\varepsilon' - \varepsilon} d\varepsilon'.
\end{eqnarray}
So we can compute the real part $\mathbb{D}(\varepsilon)$ from the imaginary part $\mathbb{I}(\varepsilon')$. The principal value ($\mathcal{P}$) avoids singularities,    and the integration variable $\varepsilon '$ is the total energy of the final state. The   threshold energy for decay $\varepsilon_M$ is the sum of all particle masses in the final states.  Then, the pole of the Green's function is
 \begin{eqnarray}
 	\varepsilon_{\rm pole} \cong M_0 +    \mathbb{D}(\varepsilon) - i\mathbb{I}(\varepsilon) ,
 \end{eqnarray}
which defines the mass and decay width of the unstable particle.
For unstable particles with decay widths are very small compared with their energy levels, we can substitute $\varepsilon$  with $ M_0 $ as an approximation. Then the pole is
\begin{eqnarray} \label{eq.modify}
	\varepsilon_{\rm pole} \cong M_0 +    \mathbb{D}(M_0) - i\mathbb{I}(M_0)  = M - i\Gamma(M_0)/2.
\end{eqnarray}
The physical mass $M$ includes a correction $\Delta M =   \mathbb{D}(M_0)$, while $\Gamma(M_0)$ is the decay width. This pole structure characterizes the resonance as a metastable state.

These equations  establish a framework for analyzing unstable composite particles using Green's functions, dispersion relations, and unitarity.
The Eq. \eqref{eq.modify} highlights the interplay between mass renormalization ($\Delta M$) and decay dynamics ($\Gamma$) in resonance physics.
 The formalism bridges bound-state Bethe-Salpeter equations with scattering theory, crucial for studying unstable states and their decays.

\begin{figure}
 	\centering
 	\includegraphics[width=1\linewidth]{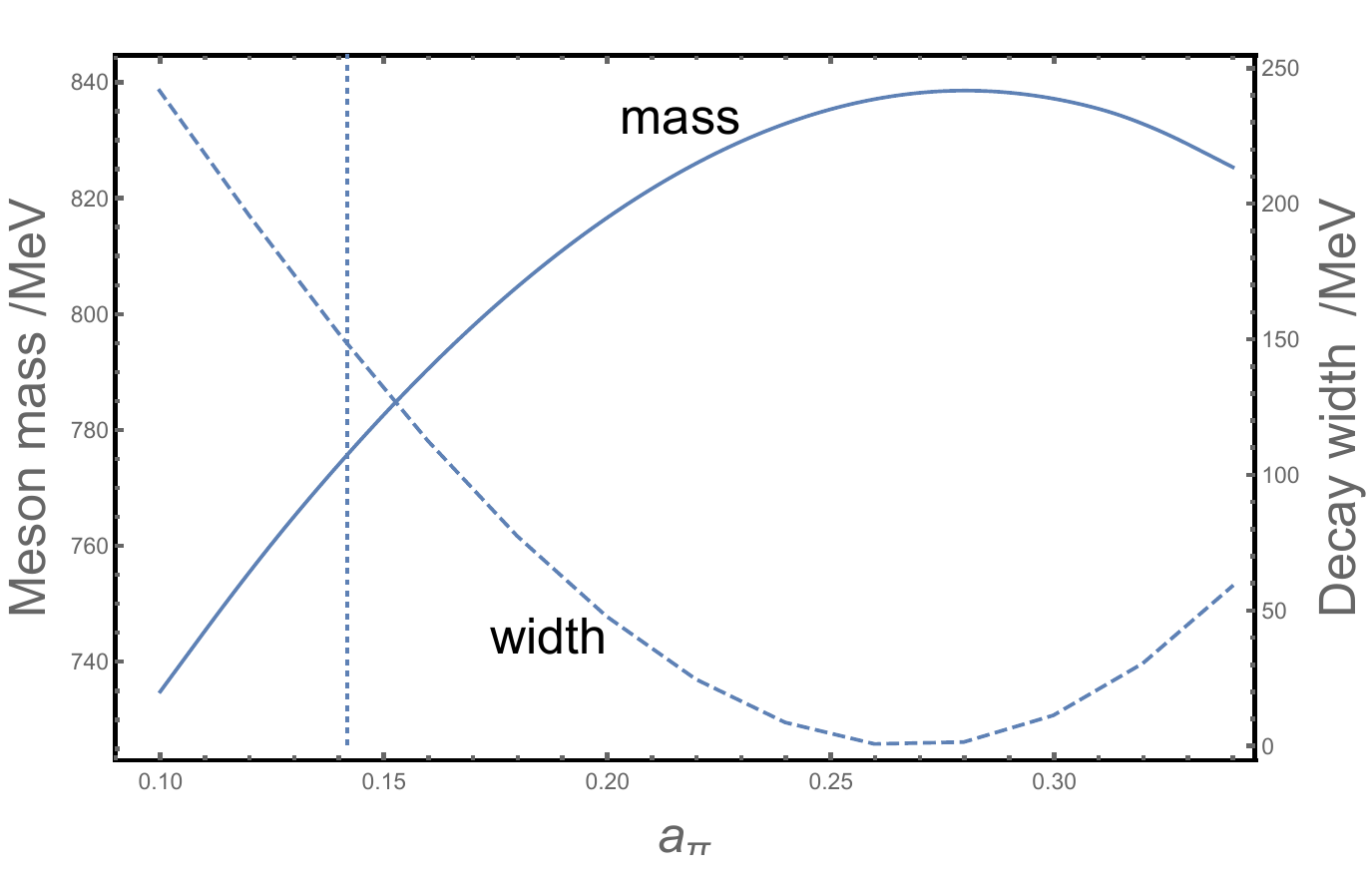}
 	\caption{  Meson mass and decay width as function of $a_\pi$.  The intersection points of the vertical line with the two curves give the experimental values.}
 	\label{fig.num}
 \end{figure}

To fit the experimental data, we adopted the following parameter settings:$\Lambda = 1050$ MeV, $ m_\pi= 139$ MeV,  $ g_{\pi-q\bar q}= 4.94$, $ g_{\rho-q\bar q}= 2.12$, and  $f_\pi = 93$ MeV which are   kept consistent with those in Ref.\cite{Polleri9611}.
Here, $G_2\Lambda^2$ and $a_\pi$ were treated as adjustable parameters. When $G_2\Lambda^2 = -57.12$ and $a_\pi$ = 0.1415, we obtained the quark mass  463.6 MeV and rho meson mass from  RPA prediction  838.6 MeV.
After corrections, the mass and decay width were calculated as   $m_\rho = 775.4$ MeV and $\Gamma= 149.3$  MeV. Comparing with the $ 1/N_c $ correction result,  the two mass correction approaches described above are  numerically consistent.   Using  $a_\pi$ as the tuning variable, the calculated mass and decay width are displayed in  Fig. \ref{fig.num}. In the plot range, while the width shows large variations over an extensive range, the mass remains confined to fluctuations around the experimental value.

%\section{Acknowledgment}

 \section{summary}
 \label{sec.sum}

 To summarize, the decay of unstable particles does not occur directly from their Bethe-Salpeter (BS) bound state. Unstable particles evolve through the BS bound state and then decay. From the BS bound state to the decay state, the mass needs to be corrected.  In this paper, taking the rho meson as an example, the decay width of the decay process to two pions is calculated.

 By selecting appropriate parameters, the article fits the experimental data of mass and decay width, and compares the results with those of the self-energy correction model under the large-$N_c$ approximation. Our calculations show that the correction of this result to the physical mass accounts for about 10\%. This may have a significant impact on more refined theoretical calculations, such as the branching ratios of other rare decays. Because in most calculations, the mass of the BS bound state is setting as the experimental mass. Here, however, the experimental mass is obtained only after correcting the BS mass.

 The article uses the relatively simple Nambu-Jona-Lasinio (NJL) model. Therefore, further, the calculation method in this paper can be extended to other model  calculations for mass correction and electro-magnetic form factors, and the analysis of other decay processes.

  	\section{Acknowledgment}
  	
This work is  supported   by the
  	Fostering Program in Disciplines Possessing Novel Features for Natural Science of Sichuan University (Grant No. 2020SCUNL209).
  	The author Q. W. Wang is extremely grateful to Professor Lü Xiaofu for his guidance in the writing of this article. The author also expresses sincere gratitude for Professor Lü Xiaofu's years of guidance and teachings.

  \appendix
  \section{The calculation of $ G $}
  \label{sec.fulu}

 The form factor appeared in Eq. \eqref{eq.width} and Eq.\eqref{eq.t} is calculated here by taking the trace of Eq. \eqref{eq.t0}  and comparing with Eq.\eqref{eq.t}. For the subsequent processing of the formula, we added a negative sign to one of the final-state momenta. Although this is not necessary, after taking the trace of the $ T $ matrix, the formula is symmetric with respect to
$p_1$ and
 $p_2$ and can be expressed as

 \begin{equation}
 	 T^\mu(k^\mu,p_1^\mu,p_2^\mu)=\int\frac{d^{4}k}{(2\pi)^{4}} \frac{8(t_1^\mu+t_2^\mu )}{m_{\pi } (s_0+s_1+s_2)} .
 \end{equation}
%neg ncnf

  with
  \begin{eqnarray}
  	s_0&=&k^2-m^2+i\varepsilon,\\
  	s_i&=&(k+p_i) ^2-m^2+i\varepsilon,\\  	
  	t_1^\mu&=&2 a_{\pi }^2 m_{\pi } k^{\mu } \left(  k\cdot  { p_1 }+ k\cdot   { p_2 }\right),
    	\end{eqnarray}

  and	\begin{eqnarray}		
  	t_2^\mu&=&   ({p_1^\mu }+{p_2^\mu }) \left(a_{\pi }^2 m_{\pi } \left(-\left(k^2+m^2\right)\right) \right. \nonumber\\
  	&+& \left. a_{\pi } m \left(-k^2+m^2+m_{\pi }^2\right)+m_{\pi } \left(k^2-m^2\right)\right) .~~~
  \end{eqnarray}
  Next, we need to express the integral of   $  T^\mu $ in terms of   $(p_1^\mu+p_2^\mu)G$ in order to obtain  $ G $.

 %  $  \text{p2}^{\mu }+\text{p1}^{\mu } $，$k_\mu$

  Note that $t_2^\mu$
   is already in the desired form. Using  $ k^2=\left(k^2-m^2\right)+m^2 $, we can express $k^2+m^2$ in terms of $s_0$. Let
  $  t_2^\mu=(  p_2 ^{\mu }+ p_1 ^{\mu } )j_2  $, then the integral corresponding to this term can be easily calculated with

  \begin{eqnarray}
  	  J_2 &=&\int\frac{d^{4}k}{(2\pi)^{4}} \dfrac{j_2}{s_0 s_1 s_2} \nonumber\\
  		&=&-a_\pi^2 m_\pi \left( I(q_\rho) +2 M(p_1,p_2) m_q^2\right) \nonumber\\
  		&+&a_\pi m_q \left(M(p_1,p_2) m_\pi^2-I(q_\rho)\right)+m_\pi I(q_\rho)
  \end{eqnarray}
 where
\begin{eqnarray}
 I(q)=\int \frac{d^{4}k}{(2\pi)^{4}} \frac{1}{(k^{2}-m^{2}+i\varepsilon)((k + q)^{2}-m^{2}+i\varepsilon)} ~~
\end{eqnarray}

and
\begin{eqnarray}
	 M(p_{1}, p_{2})&=&\int \frac{d^{4}k}{(2\pi)^{4}} \frac{1}{s_0s_1s_2}.
\end{eqnarray}
 The integrals  $ I $ and $ M $
can be easily calculated using the Feynman parametrization method.

 When the final-state pions are on shell, the integral of the term corresponding to
$t_1^\mu$ is expressed as
  \begin{eqnarray}
  	(p_1^\mu+p_2^\mu) J_1  &=&2 a_{\pi }^2 m_{\pi } \int \frac{d^{4}k}{(2\pi)^{4}}\dfrac{2k^{\mu } \left( k \cdot p_1
  	 	+k\cdot p_2 \right)}{ s_0s_1s_2 }\nonumber \\
   	&=&2 a_{\pi }^2 m_{\pi }(a_1+a_2 +b+c ),
  \end{eqnarray}
  with
  \begin{eqnarray}
  	a_i&=& \int \frac{d^{4}k}{(2\pi)^{4}} \frac{k^{\mu }}{ s_0s_i},   \\
  	b&=&\int\frac{d^{4}k}{(2\pi)^{4}}\frac{2k^{\mu }}{s_1s_2}, \\
  	c&=&-  (p_1^2 + p_2^2) \int\frac{d^{4}k}{(2\pi)^{4}}\frac{k^{\mu }}{  s_0s_1s_2 }.
  \end{eqnarray}
   After some cumbersome integrations, we can obtain
  \begin{eqnarray}
 	a_i&=&-\frac{p_i^{\mu } }{2} I(p_i),    \\
 	b&=& (p_1^\mu+p_2^\mu  )  I (p_1-p_2) , \\
 	c&=&- (p_1^2+p_2^2) (  M_1(p1,p2) p_1^\mu+M_1(p2,p1)p_2^\mu ),~~~~~~
 \end{eqnarray}
with
\begin{eqnarray}
	M_{1}(p_{1}, p_{2})=\frac{1 }{2((p_{1} \cdot  p_{2})^{2}-p_{1}^{2}p_{2}^{2})}
	( p_{1} \cdot p_{2}I(p_{1})~~ \nonumber\\
	-p_{2}^{2}I(p_{2})
	+  (p_{2}^{2}-p_{1} \cdot p_{2})I(p_{1}-p_{2})\nonumber\\
	+p_{2}^{2}(p_{1}^{2}-p_{1} \cdot p_{2})M(p_{1}, p_{2})) . ~~
\end{eqnarray}
The calculation of $ M_1 $  can make use of the assumption
\begin{eqnarray}
 p_{1}^{\mu}M_{1}(p_{1}, p_{2})+p_{2}^{\mu}M_{1}(p_{2}, p_{1}) = \int \frac{d^{4}k}{(2\pi)^{4}} \frac{k^{\mu}}{s_0s_1s_2} .   ~~
\end{eqnarray}
The calculated
$  J_1 $ is written as
  \begin{eqnarray}
   J_1=   -\frac{1}{2}I(p_i) + I (p_1-p_2)
    - ( p_1^2+p_2^2)  M_1 (p1,p2 ).~~~
  \end{eqnarray}
Finally, the function $ G $ can be written in terms of $ J_1 $ and $ J_2 $, with
  \begin{eqnarray}
  G(\rho_m^2,\pi_m^2) =\frac{8 }{m_\pi}	( 2 a_\pi^2 m_\pi J_1+J_2 ).
  \end{eqnarray}

 % \[I_{1}=\int \frac{d^{4}k}{(2\pi)^{4}} \frac{1}{k^{2}-m^{2}+i\varepsilon}\]

%  \begin{eqnarray}
%  	& & L(p_{1}, p_{2}, p_{3}) \\
%  	&=&\int \frac{d^{4}k}{(2\pi)^{4}} \frac{1}{(k^{2}-m^{2}+i\varepsilon)(k_{1}^{2}-m^{2}+i\varepsilon)(k_{2}^{2}-m^{2}+i\varepsilon)(k_{3}^{2}-m^{2}+i\varepsilon)}
%  \end{eqnarray}

%	\appendix
%	\section{title}
\bibliographystyle{plain}
\bibliography{rho25}

% \bibliographystyle{apalike}
%	\bibliography{rho25}
%	\end{CJK}
\end{document}